\documentclass[superscriptaddress,reprint,secnumarabic,amssymb, nobibnotes, apl]{revtex4-1}

\newcommand{\im}{\operatorname{Im}}

\DeclareMathAlphabet{\mathpzc}{OT1}{pzc}{m}{it}

\usepackage{graphics}
\usepackage{epsfig}
\usepackage{epstopdf}
\usepackage{subfigure}
\usepackage{amsmath}
\usepackage{subfigure}
\usepackage{float}
\usepackage{enumerate}
\usepackage{array}
\usepackage{pifont}
\usepackage[11pt]{moresize}

\usepackage{datetime}

\usepackage{mathrsfs}
\newcolumntype{C}[1]{>{\centering\let\newline\\\arraybackslash\hspace{0pt}}m{#1}}
\newcolumntype{N}{@{}m{0pt}@{}}

\begin{document}

	
	\title{Universal Zero Conductivity Condition for Optical Absorption}
	
	
	\affiliation{Department of Electrical Engineering University of Alberta, Edmonton, AB T6C 2V4, Canada}
	
	\author{Yu Guo}
	\affiliation{Department of Electrical Engineering University of Alberta, Edmonton, AB T6C 2V4, Canada}
	\author{Sarang Pendharker}
	\affiliation{Department of Electrical Engineering University of Alberta, Edmonton, AB T6C 2V4, Canada}
	
	\author{Zubin Jacob}
	\affiliation{Department of Electrical Engineering University of Alberta, Edmonton, AB T6C 2V4, Canada}
	\affiliation{Birck Nanotechnology Center, School of Electrical and Computer Engineering, Purdue University, West Lafayette, IN 47906, USA }
	
	
	
	\begin{abstract}
		 Harnessing information and energy from light within a nanoscale mode volume is a fundamental challenge for nanophotonic applications ranging from solar photovoltaics to single photon detectors. Here, we show the existence of a universal condition in materials that sheds light on fundamental limits of electromagnetic to matter energy conversion (transduction). We show that the upper limit of absorption rate (transduction rate) in any nanoscale absorber converting light to matter degrees of freedom is revealed by the zero of optical conductivity at complex frequencies  ($\sigma(\omega^\prime + i\omega^{\prime\prime})= 0$). 
We trace the origin of this universal zero conductivity condition to causality requirements on any passive linear response function and propose an experiment of absorption resonances using plasmonic nanoparticles to experimentally verify this universal zero conductivity condition.  Our work is widely applicable to linear systems across the electromagnetic spectrum and allows for systematic optimization of optical absorption in single photon detectors, solar cells, Coherent Perfect Absorbers and SPASERS.  

	\end{abstract}
	\maketitle
Fundamental limits to light trapping, optical absorption and energy conversion directly depend on the dielectric response of materials as well as the geometric device design and are of vital importance for optimizing practical photonic devices \cite{Raman2013,yablonovitch1982statistical,khurgin2011scaling,wang2006general,miller2016fundamental,miller2014fundamental,yu2010fundamental,min2010enhancement,sersic2011magnetoelectric,miller2015shape}. The material response (complex refractive index or optical conductivity) is a macroscopic quantity which contains information on varied microscopic excitations such as electron-hole pairs in semiconductors, free electron oscillations in metals and vibrational phonon modes in polar dielectrics. Thus material constraints, which are independent of device geometry, can guide design and optimization of photon transduction for devices such as solar cells and photodetectors. 

Photonic device response, in the form of scatterers, waveguides or resonators, is characterized by the scattering matrix of electromagnetic waves \cite{stone2001wave}. The excited resonant modes in passive photonic devices decay with time which signifies a complex resonant frequency ($\omega_{res}=\omega^\prime + i\omega^{\prime\prime}$).  The resonant modal energy has to be slowly dissipated as heat or radiated into vacuum imposing the condition $\omega^{\prime\prime} <0$ (for $e^{-j\omega t}$ convention). Here, we show that exploring the material response into this complex frequency regime of $\omega^{\prime\prime} <0$ provides fundamental insight on the absorption and scattering response of all photonic devices.

The subject of this paper is the observation, which has been surprisingly overlooked till date, that the optical conductivity (or equivalently the imaginary part of the dielectric constant) of any material has a zero in the lower-half of the complex frequency plane $\omega^{\prime\prime} <0$. The significance of our result becomes clear on contrasting with  Landau's insight \cite{landau2013electrodynamics} that the optical conductivity cannot be zero anywhere on the real frequency axis or the upper half of the complex frequency plane. We explain the physical significance of this universal zero conductivity condition as the upper limit to the absorption rate (transduction rate) for a given material. We show that this upper limit occurs for a plasmon resonance of a metallic nanoparticle. We also provide an appealing physical picture of why such a zero conductivity condition should be universal for all passive materials. We finally propose experiments to verify the existence of this universal condition using plasmonic nanoparticles and show this condition sheds light on perfomance characteristics of SPASERS \cite{stockman2010spaser} and Coherent Perfect Absorbers (CPAs) \cite{chong2010coherent}.
Recently, important progress has been made in identifying the upper limit to material modal loss rate in terms of Drude damping coefficients \cite{Raman2013}. There has also been work related to limits of absorption in ray optics \cite{yablonovitch1982statistical}, plasmonic resonators \cite{khurgin2011scaling,wang2006general}, wave optics \cite{miller2016fundamental,miller2014fundamental}, thin films \cite{min2010enhancement}, solar cells \cite{yu2010fundamental}, upper limits of magneto-optic cross-coupling \cite{sersic2011magnetoelectric} and limits of near-field radiative heat transfer \cite{miller2015shape}. However, the zero conductivity condition and its universality has not been pointed out before. Our result is also applicable to all passive linear response functions thus providing a generalization of previous results.

The material response function at optical frequencies is given by the complex dielectric permittivity ($\epsilon(\omega)=\epsilon^{\prime}(\omega) + i\epsilon^{\prime\prime}(\omega)$) and magnetic permeability ($\mu=1$). We identify the frequency dependent optical conductivity with the imaginary part of the spectrally dispersive dielectric constant $\sigma(\omega)=\frac{Im(\epsilon(\omega))}{\omega}$. In \cite{landau2013electrodynamics}, Landau shows that the imaginary part of dielectric response in any passive material has to be greater than zero for all positive real frequencies. Furthermore, it is also shown that the conductivity cannot be zero in the upper-half of the complex frequency plane. Our central result is the subtle observation that the optical conductivity can be zero in the lower half of the complex plane i.e.
\begin{equation}
\sigma(\omega^{\prime} + i\omega^{\prime\prime})= 0
\end{equation}

\begin{figure}
\centering
\subfigure{\includegraphics[width=1\linewidth]{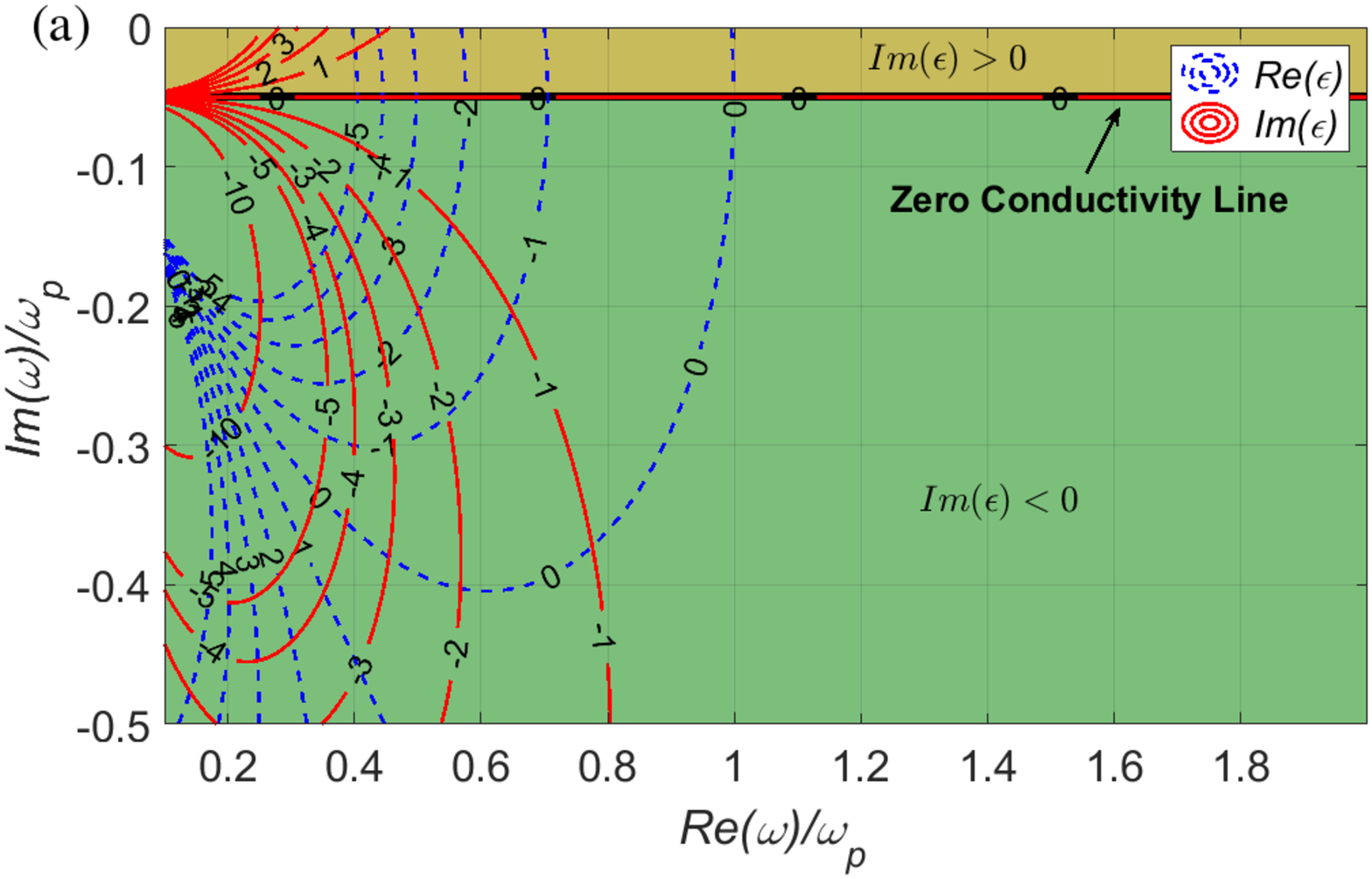}}
\subfigure{\includegraphics[width=1\linewidth]{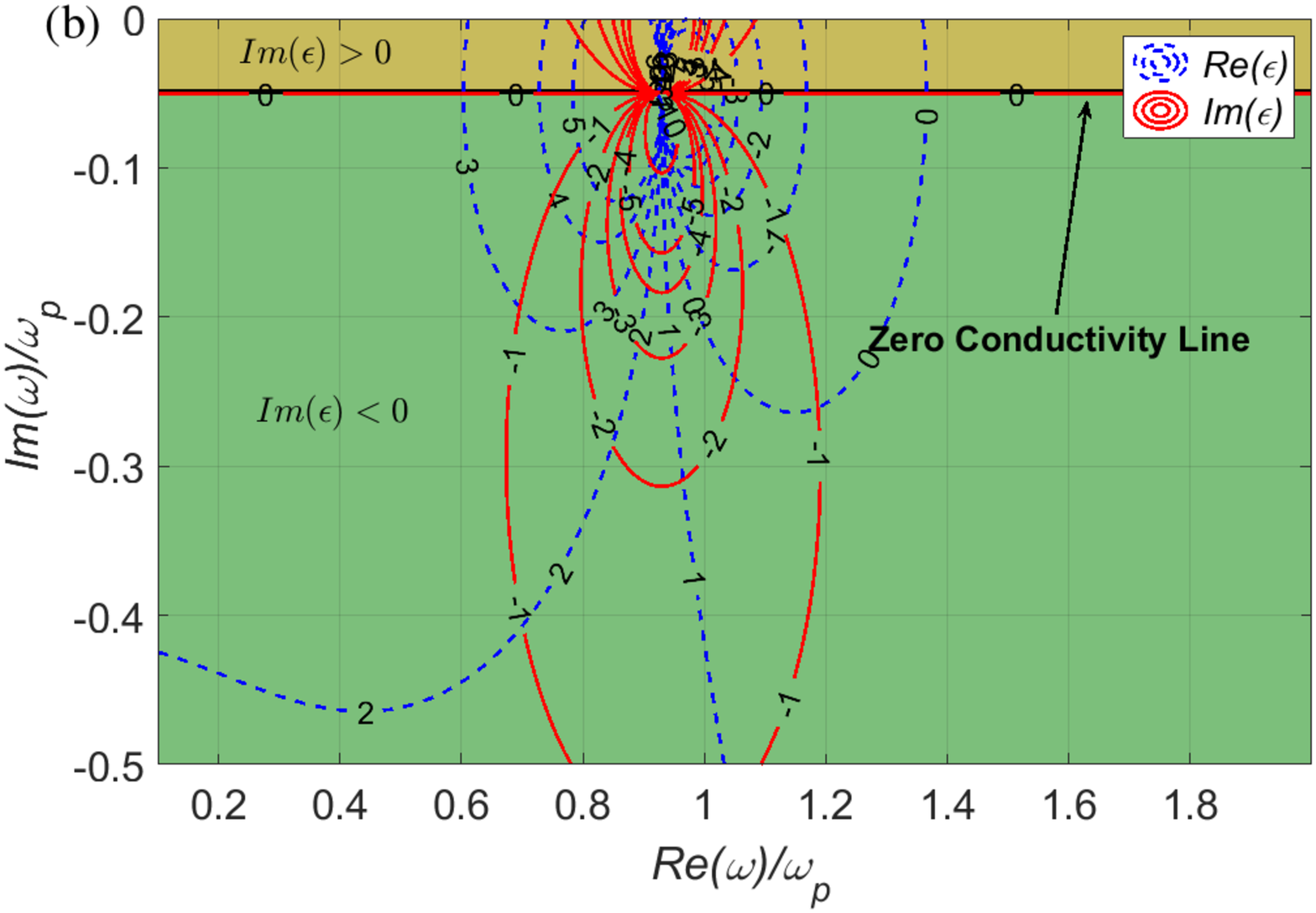}}
\caption{Dielectric response of an optical material using Drude (a) and Lorentz (b) model in the complex $\omega$ plane. Blue curves represent constant real part of the dielectric permittivity ($\epsilon^\prime$) and red curves represent constant imaginary part ($\epsilon^{\prime\prime}$). The green background represents the region of negative conductivity and the yellow region represents positive conductivity. Our central result is that the transition between positive and negative conductivity happens on the zero conductivity line which is universal for all materials. For the Drude and Lorentz case we have $Im(\omega)=-\Gamma/2$ as the zero conductivity line.}
\label{Drude_dieletric_complex_omega2}
\end{figure}

We now identify this zero conductivity condition in the familiar Drude and Lorentz optical response functions. Fig.~\ref{Drude_dieletric_complex_omega2} shows the dielectric response of Drude metal ($ \epsilon_r=1-\omega_{p}^2/(\omega^2+i\Gamma\omega)$) and Lorentz metal ($ \epsilon_r=1+\omega_{p}^2/\left(\left(\omega_0^2-\omega^2\right)-i\Gamma\omega\right)$) in the complex frequency plane $\omega= \omega^\prime +i\omega^{\prime\prime}$. Note, the key difference from the dielectric response function conventionally evaluated only at real frequencies. The blue and red curves represent real ($\epsilon^\prime$) and imaginary ($\epsilon^{\prime\prime}$) components of the dielectric response respectively. The sign of $\epsilon^{\prime\prime}$ in the $\omega^\prime +i\omega^{\prime\prime}$ plane is represented by the background color. The green region represents $\epsilon^{\prime\prime}<0$, while yellow region represent $\epsilon^{\prime\prime}>0$. It can be seen that for  both Drude and Lorentz metal, $\epsilon^{\prime\prime}<0$ and $\epsilon^{\prime\prime}>0$ regions are separated by a line on which the conductivity is zero. We will refer to this as the zero conductivity curve. It can be shown analytically that for Drude and Lorentz metals, the zero conductive curve is a straight line in the complex frequency plane given by $\omega^{\prime\prime}=-\Gamma/2$, where $\Gamma$ is the damping frequency. 
\begin{figure}
\centering
\subfigure{\includegraphics[width=1\linewidth]{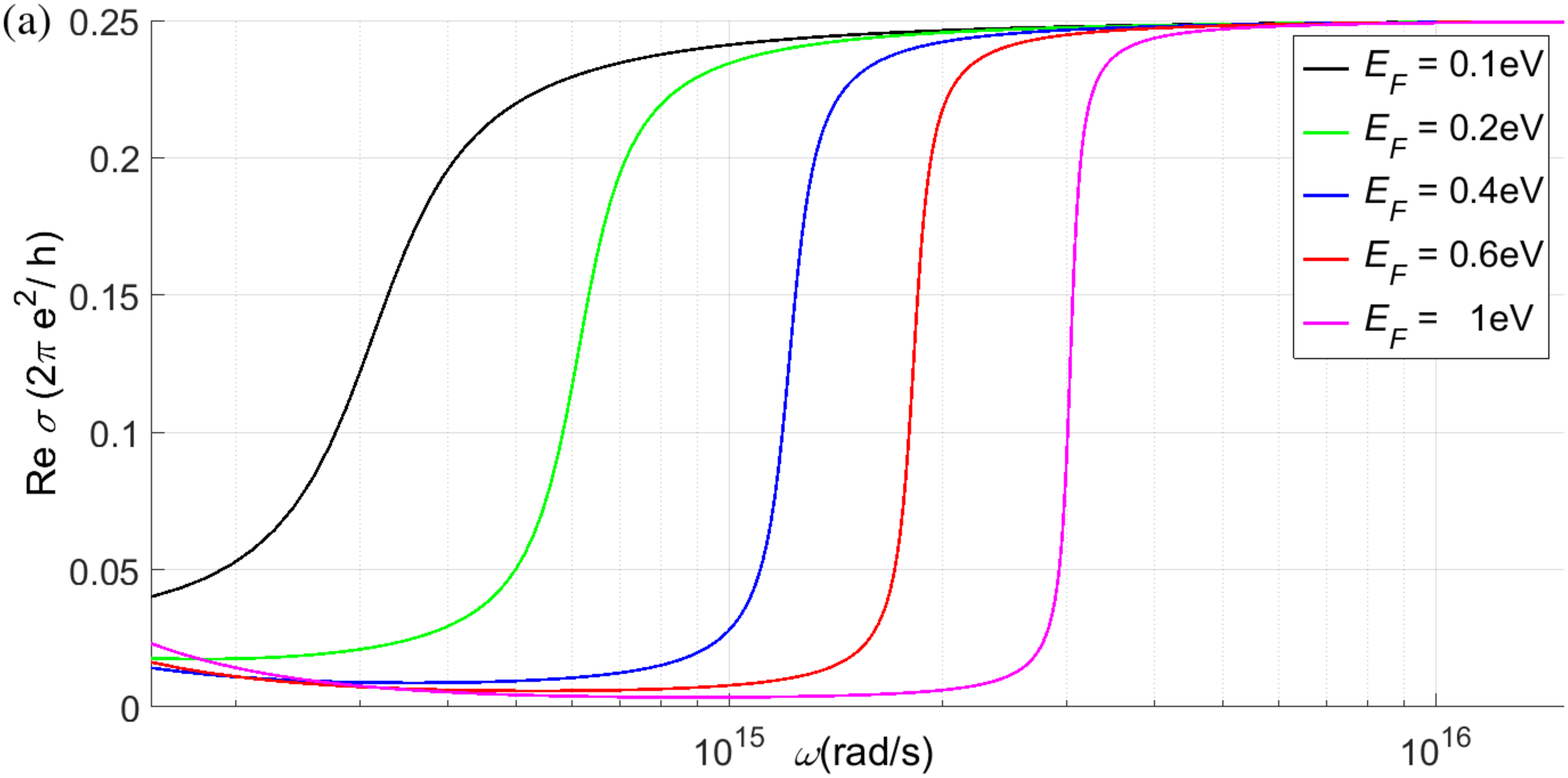}}
\subfigure{\includegraphics[width=1\linewidth]{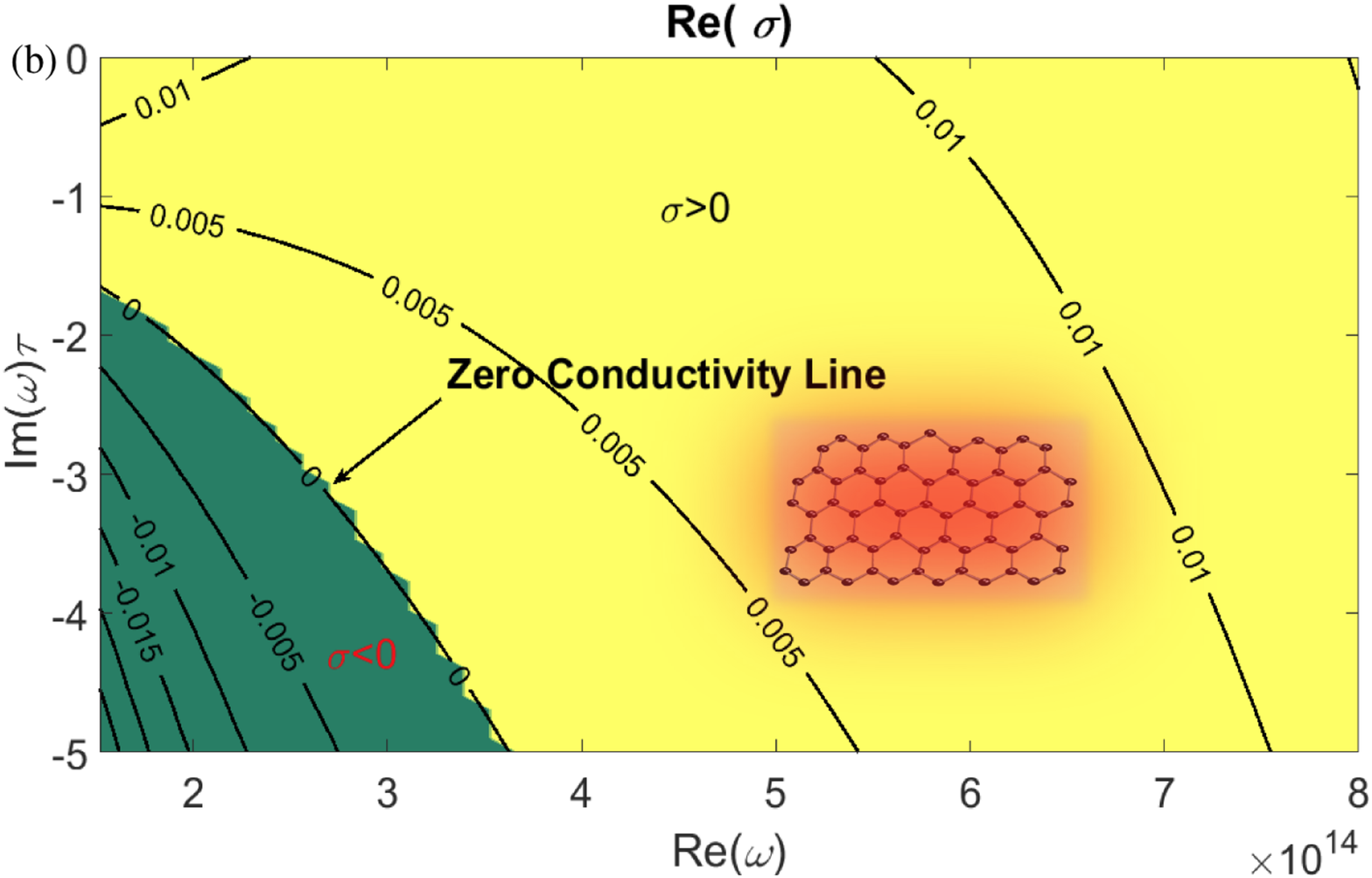}}
\caption{Conductivity of graphene as (a) function of real frequencies and (b) in the complex frequency plane. The green background in (b) represents the negative conductivity region and the yellow background represents the positive conductivity region, which are separated by the zero conductivity curve. (b) is computed for Fermi Energy $E_F=0.4$eV. Unlike a Drude or Lorentz metal the zero conductivity line is not a straight curve, nevertheless, it exists even in a 2D material like graphene.}
\label{graphene_Re_conductivity_T=300_closed_form_tau=1e-12}
\end{figure}

The zero conductivity curve is not unique to Drude-Lorentz materials. Fig.~\ref{graphene_Re_conductivity_T=300_closed_form_tau=1e-12} shows the conductivity of graphene \cite{jablan2009plasmonics,thongrattanasiri2012complete} for real and complex frequencies. Even for a 2D material like graphene, there exits a zero conductivity curve separating the $\sigma>0$ region and region with $\sigma<0$ in the complex frequency plane. A universal behavior is seen in all the response functions that the zero conductivity line separates the regions where the material response to complex frequencies is active or passive.  

We will now apply the concept of the zero conductivity condition to shed light on the limits of the absorption rate in nanoscale structures. The absorption and scattering of light from cylinders and spheres is described by Mie theory and is strongly enhanced near resonances \cite{bohren2008absorption}. These resonances are captured by the poles of the scattering matrix (reflection coefficient) in the complex frequency plane.  In general, the poles of the scattering matrix can lie anywhere in the lower half of the complex frequency plane but the maximal absorption (transduction) rate where electromagnetic energy is dissipated as heat (matter degrees of freedom) is limited by the zero conductivity line (see Supp. Info.). This arises since the dielectric response function in the complex frequency plane becomes active beyond the zero conductivity condition which is prohibited specifically for energy transduction. 

\begin{figure}
\centering
\includegraphics[width=1\linewidth]{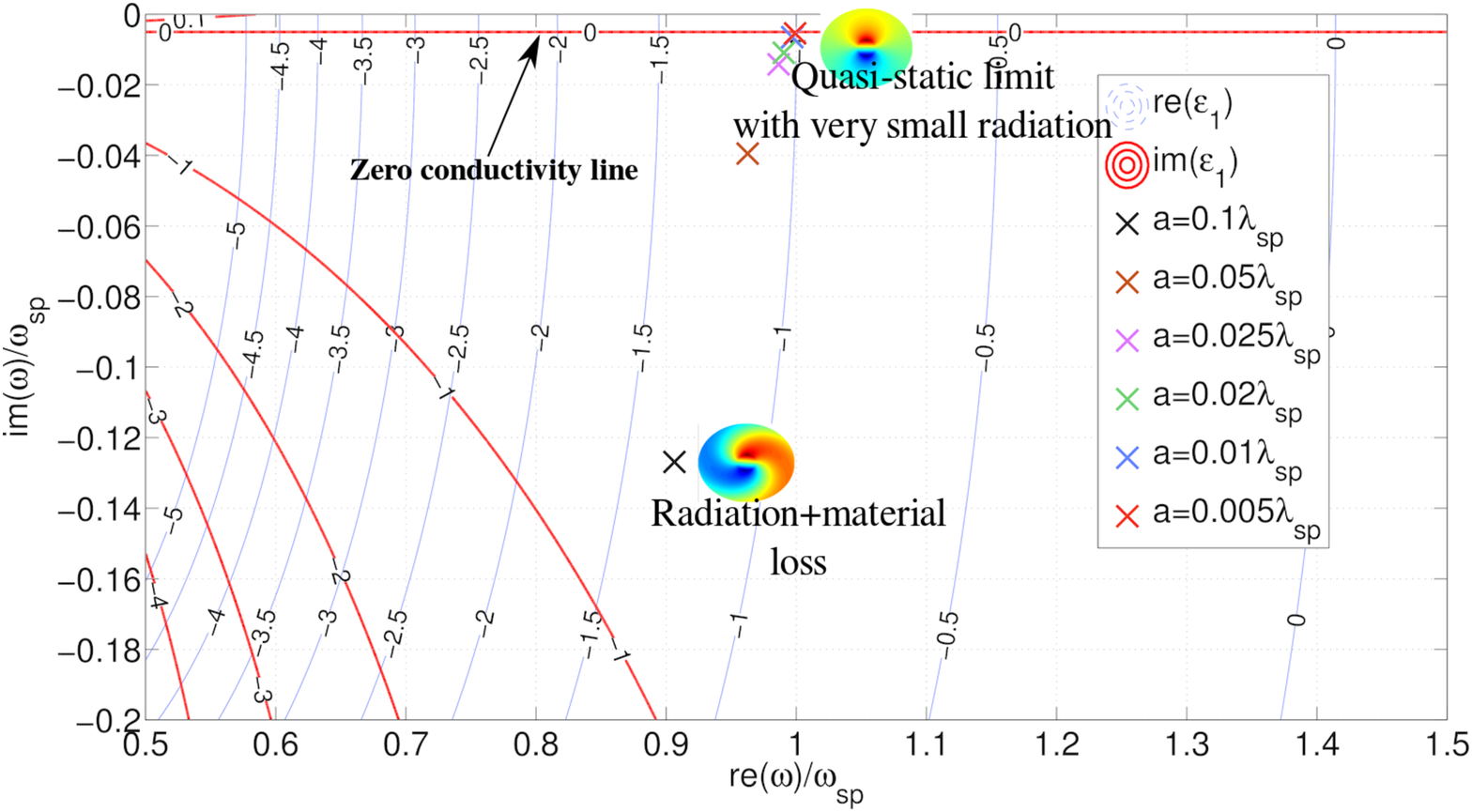}
\caption{Eigenmode solutions of a plasmonic nanowire plotted in the complex frequency plane. The real part of the solution corresponds to the resonant frequency, while the imaginary component governs the total decay rate. The loss rate, and consequently the imaginary component of the resonant frequency, has contributions from modal material loss as well as radiation loss. As the radius of nanowire decreases the eigenfrequency approaches the zero conductivity line of $\epsilon^{\prime\prime}=0$ and the absorption (transduction) rate achieves its maximum value. The inset shows field profiles for $a=0.1\lambda_p$ and $a=0.005\lambda_p$. For the large radius the field has a major radiating component, for small radius the field approaches a quasi-static limit and the net loss is dominated by modal material loss. Here a Drude metal with $\Gamma=0.01\omega_{p}$ is considered. The blue contours represent constant $\epsilon^\prime$ and the red contours represent constant $\epsilon^{\prime\prime}$. The frequency is normalized with the surface plasmon frequency $\omega_{sp}$, where $\sqrt{2}\omega_{sp}=\omega_p$.}
\label{drude_cylinder_vary_a_quasi_static_limit}
\end{figure}

Interestingly, for the case of metals (eg: gold, silver), the plasmonic resonances consisting of light oscillating with free electrons, the poles lie on the zero conductivity line. This occurs in the quasi-static limit and the absorption (transduction) rate achieves its maximum value.  Intuitively, this upper limit is expected since electric energy in the fields in the quasi-static limit cannot be converted into heat faster than half the electron collision rate in the plasmonic medium. Note, the magnetic energy is negligible in the quasi-static limit \cite{khurgin2011scaling}. Fig.~\ref{drude_cylinder_vary_a_quasi_static_limit} shows the poles of the scattering matrix in the complex frequency plane for a nanowire made of Drude metal. The resonant frequency of the mode is the real component ($\omega^\prime$) of the solution  while the imaginary part ($\omega^{\prime\prime}$) represents the total modal decay rate. The total modal decay rate $\omega^{\prime\prime}$ has contributions from two effects: the radiative decay of energy from the mode and heat dissipation, also called the modal material loss \cite{Raman2013}.  It can be seen in Fig~\ref{drude_cylinder_vary_a_quasi_static_limit}, that the total modal loss rate is greater than $\Gamma/2$ in general. However, in the quasi-static limit when the particle size is much smaller than the incident wavelength, there exists no radiation loss and the loss is dominated by the modal material loss. In this case, the eigenmode approaches the zero conductivity curve in the complex frequency plane ($\omega^{\prime\prime}=-\Gamma/2$). In this limit, the pole lies on the zero conductivity curve where modal material loss (energy dissipated as heat) dominates over the radiative loss of energy. The inset shows the field profiles for two different cases when the total decay rate of the eigenmode is dominated by radiation-loss ($a=0.1\lambda_{sp}$)  or by material-loss ($a=0.005\lambda_{sp}$). 

\begin{figure}
\centering
\includegraphics[width=1\linewidth]{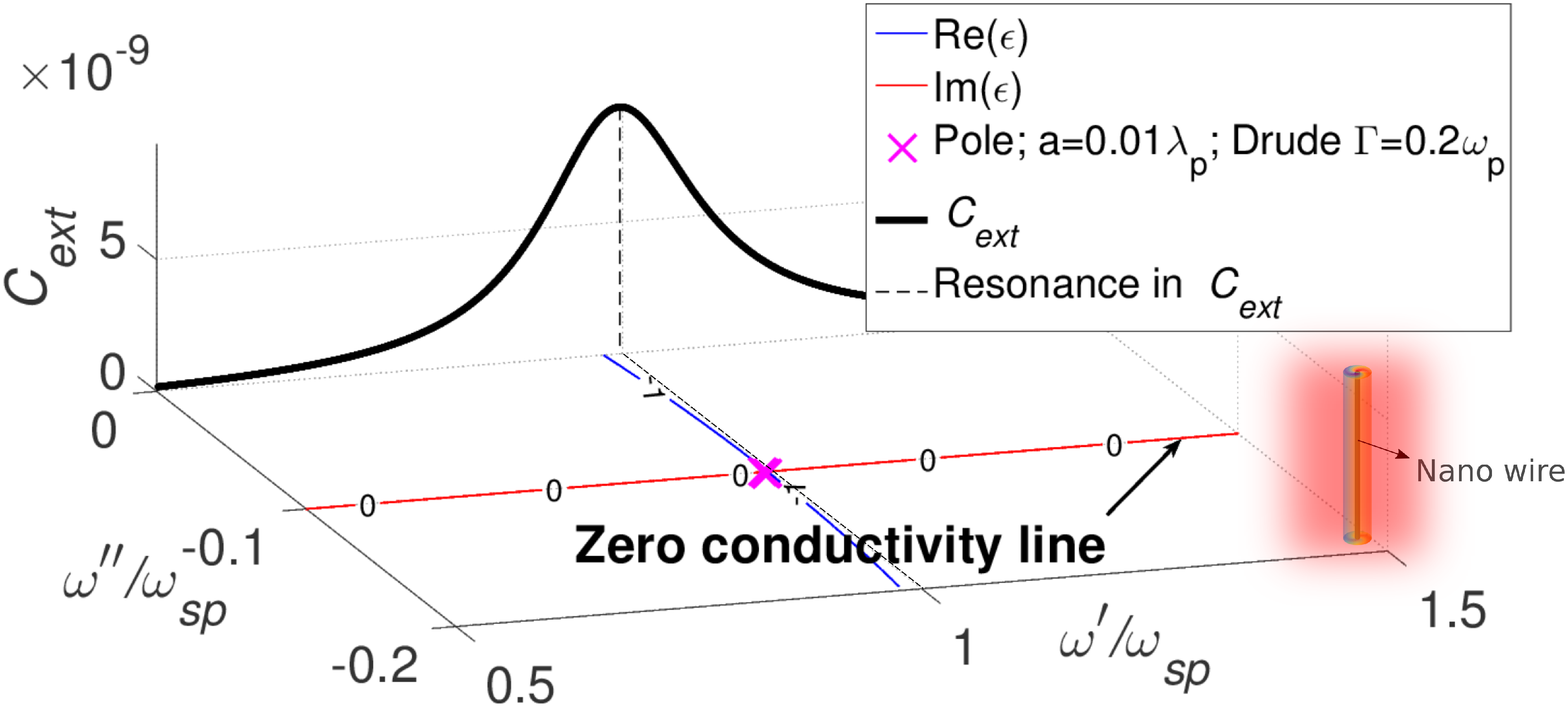}
\caption{Eigenmode of nanowire ( Drude metal $\Gamma=0.2\omega_{p}$) in the complex frequency plane and the corresponding extinction cross-section in the quasi-static limit with radius $a=0.005\lambda_{p}$. The resonance peak in the frequency response of a nanowire to plane wave excitation depends on the location of the eigenmode in the complex frequency plane. The resonant frequency is the value of $\omega^\prime$ at which the contour $\epsilon^\prime=-1$ intersects the zero conductivity line $\epsilon^{\prime\prime}=0$. Our theory can be verified through the peak and quality factor of resonance which is directly proportional to $\omega^\prime$ and $\frac{\omega^\prime}{\omega^{\prime\prime}}$ respectively.}
\label{Scattering+Eigen_value_poles_D_a_0pt01lambda0_Gamma_0pt2}
\end{figure}

The zero conductivity mode also gives fundamental insight into the extremely well-known plasmonic resonance condition. The spectral peak in the absorption and scattering response of the plasmonic nanowire is usually evaluated using $\epsilon^\prime=-1$. However, as described above, the plasmonic pole in the complex frequency plane also lies on the zero conductivity curve. Thus the exact spectral location of the eigenmode is on the intersection of the  zero conductivity curve and $\epsilon=-1$ contour in the complex frequency plane. This is depicted in Fig.~\ref{Scattering+Eigen_value_poles_D_a_0pt01lambda0_Gamma_0pt2}.  It can be seen that the resonance frequency in the extinction cross-section ($C_{ext}$) coincides with real component of the eigenfrequency of the plasmonic mode on the zero conductivity curve, and not with $\epsilon^\prime=-1$ on the real axis (Fig.~\ref{Scattering+Eigen_value_poles_D_a_0pt01lambda0_Gamma_0pt2}). 

We now propose an experiment to confirm the concept of the zero-conductivity condition. This can be done through experimentally measured plasmonic resonances in nanospheres or nanowires made of different plasmonic materials \cite{bohren2008absorption,naik2012titanium,oldenburg1998nanoengineering} (eg: gold, silver and titanium nitride).  The resonant frequency will give the real component of the eigenfrequency ($\omega^\prime$) in the the complex frequency plane, while the imaginary part ($-\omega^{\prime\prime}$) will be computed from the measured Q-factor of the resonance, since $Q \propto \frac{\omega^\prime}{\omega^{\prime\prime}}$. Using ellipsometry, the dielectric response for the same metals  can be empirically estimated as a function of real frequency $\omega$. Subsequently, we can use the analytic continuity of the dielectric response to evaluate the dielectric response specifically at complex frequencies. We predict the imaginary component of the dielectric response at the complex resonance frequency to be close to zero ($\epsilon^{\prime\prime}(\omega^\prime+i\omega^{\prime\prime})=0$) which is the zero conductivity condition. We expect deviations due to the role of the interband transitions but it can be isolated using careful characterization. 

\begin{figure}
	\centering
	\includegraphics[width=0.45\linewidth]{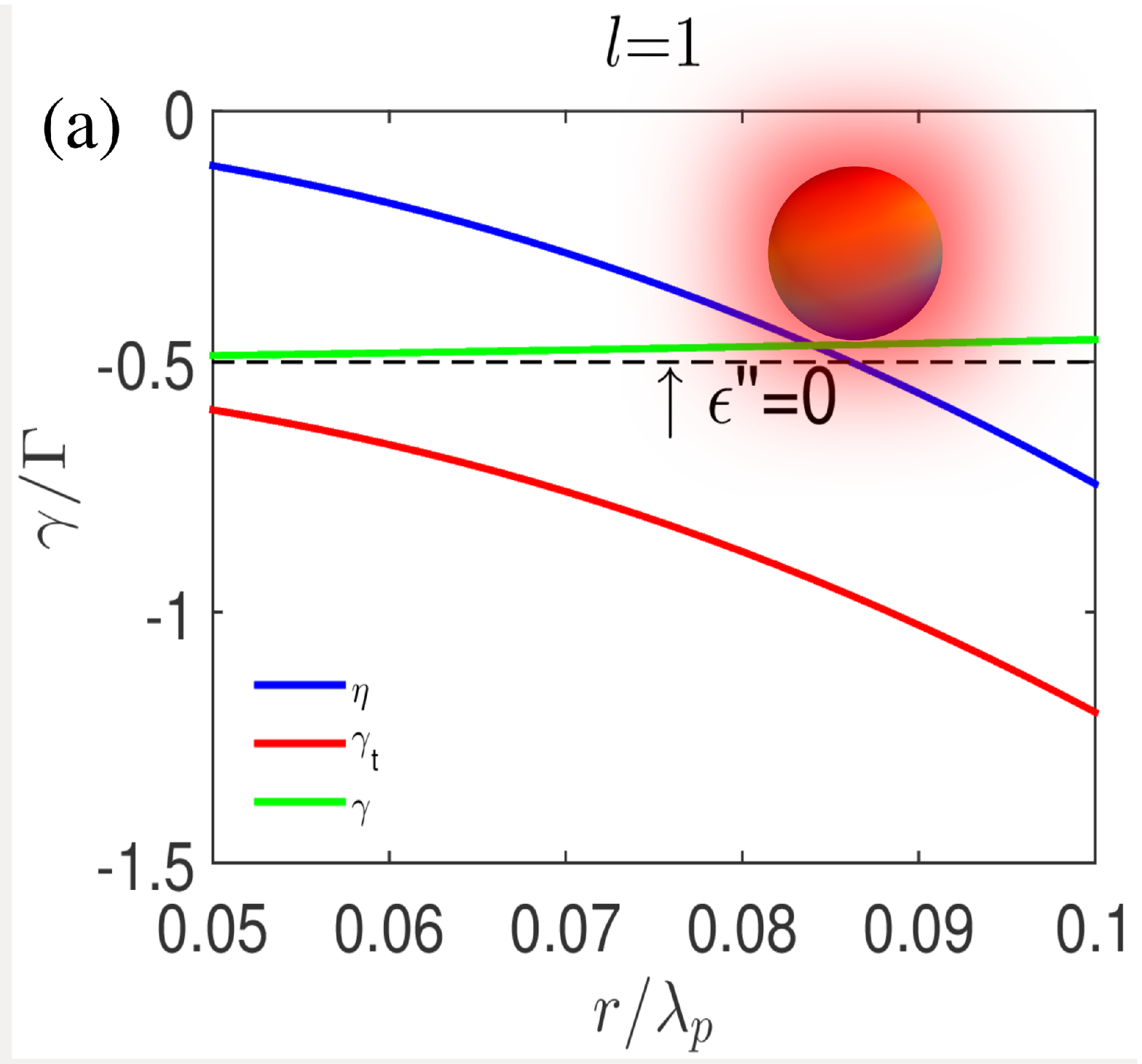}
	\includegraphics[width=0.45\linewidth]{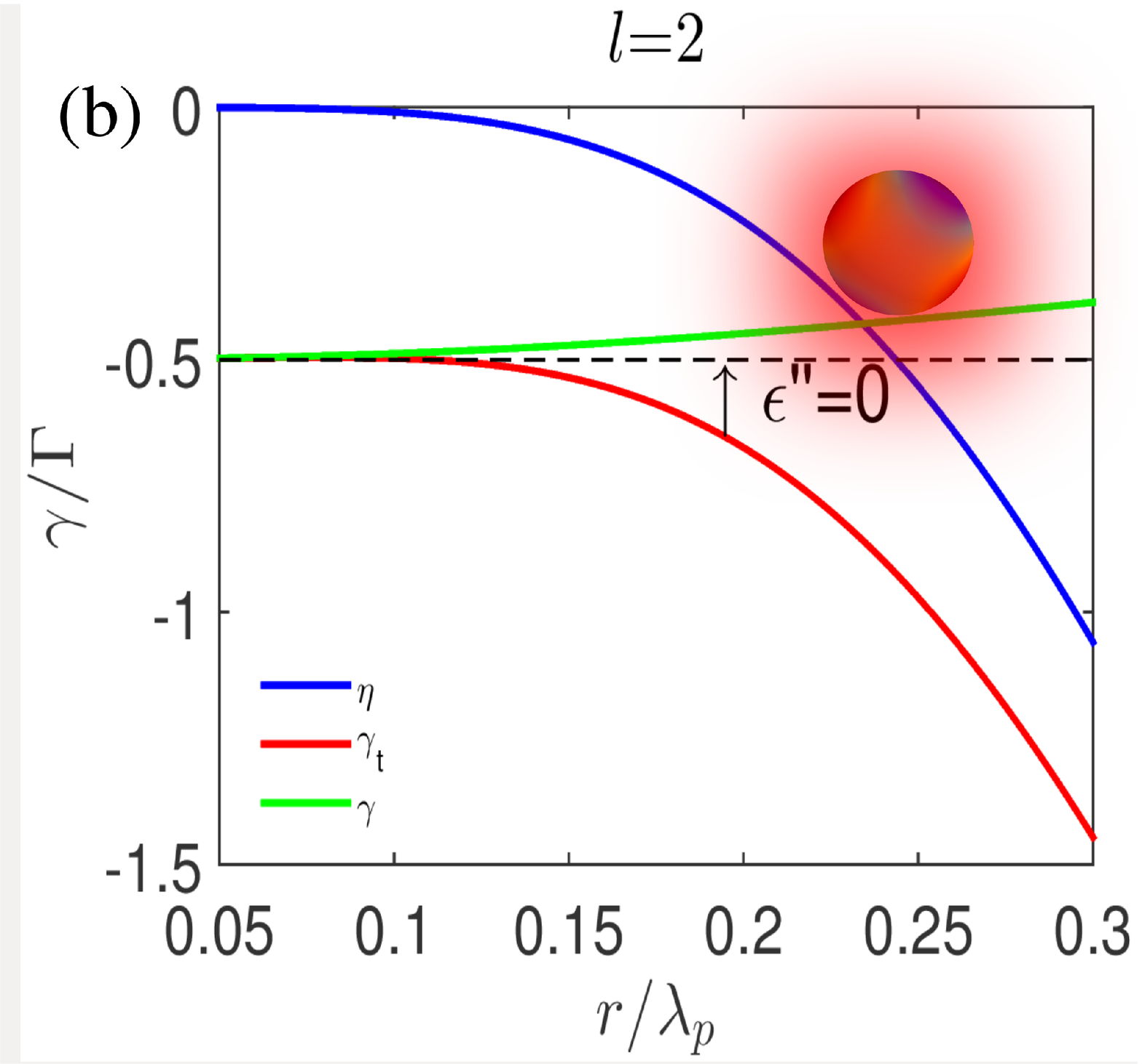}
	\caption{\label{fig:sphere_wi_L1} Imaginary part of eigen-frequencies of the localized surface plasmon modes of nanospheres (a) for $l=1$ and (b) $l=2$. Here $\eta$ the radiation loss i.e. the modal loss when the material is lossless. $\gamma_t$ is the modal loss when the material is lossy, which includes the radiation loss and the material loss. The material loss is approximated by $\gamma=\gamma_t-\eta$. The dash black line denotes the condition $\epsilon''=0$. Thus in both cases the material loss rate $\gamma$ satisfies the condition that $\epsilon''(\omega'+i\gamma)>0$ and approaches the zero conductivity condition in the limit of small radius spheres (quasi-static limit).}
\end{figure}

Next, we calculate separately the radiation loss rate and material loss rate for plasmonic nanospheres. This shows that the upper limit of the modal absorption rate is achieved when the pole approaches the zero-conductivity curve. First, we compute the complex eigenfrequencies by finding poles of the Mie scattering coefficients of $l$-th order TM modes \cite{kong1975theory} . 
This gives us the total decay rate of the plasmonic mode denoted by  $\gamma_t$. To separate the absorption rate (transduction rate) denoted by $\gamma$, at which electromagnetic energy is dissipated as heat, we calculate the eigenfrequencies for two cases i) lossless ii) with loss. In the supplementary information, we have shown detailed justification of our approach. 
The blue curves in Fig.\ref{fig:sphere_wi_L1}(a)-(b) represent the net loss rate in a sphere with zero material loss ($\Gamma =0$), as a function of normalized sphere radius $r/\lambda_p$, for the first($l=1$) and second ($l=2$) order modes. Since there is no material loss, the net loss rate corresponds to the radiation loss rate $\eta$ in the nano-sphere.  The red curve represents the net loss rate $\gamma_t$ in a lossy Drude sphere with $\gamma=0.01\omega_p$. We then approximate the material loss rate by $\gamma=\gamma_t-\eta$, shown by green curve in the figure. This approximation is justified since the excited mode profile and the resonant frequency $\omega^\prime$ for lossy and loss-less are almost identical (see supplementary information Fig.~\ref{fig:sphere_wi_L1_supp} ). It can be seen that the modal material loss rate is confined in the region $\epsilon''(\omega'+i\omega'')>0$ and approaches the upper limit of $\Gamma/2$ when the pole approaches the zero-conductivity curve $\epsilon''(\omega'+i\omega'')=0$ (shown in black dashed curve). A detailed analysis of the optimal absorption cross-section of the nanospheres is provided in the supp. info.. 
	

Our work also sheds light on SPASERS \cite{stockman2010spaser} and CPAs \cite{chong2010coherent}. Spasing (equivalent to lasing) is achieved when the plasmonic pole of the scattering matrix is pulled up the real frequency axis by the action of the gain medium ($Im(\epsilon)<0$). As shown before, for nanoscale plasmonic resonators, the plasmon pole lies on the zero conductivity line. This is the case when there is maximum dissipation of electromagnetic energy as heat, thus the gain medium has to overcome the maximum amount of material loss in the system for SPASING. Our work suggests the need for core-shell structures where the plasmon pole is shifted from the zero conductivity line as close to the real axis as possible. Similarly, for CPAs (supp. info. Fig. 1), an engineered amount of loss brings a zero of the scattering matrix to the real axis for perfect absorption. This is equivalent to the critical coupling condition when modal material loss rate is balanced by the radiation loss rate. However, the existing theory does not shed light on the optimal absorption rate. Our work suggests an interplay between the zero conductivity condition and location of CPA poles and zeros to optimize both absorption cross-section and absorption rate (supp. info. Fig.~\ref{fig:sphere_wi_L1_supp}). 

To summarize, in this paper we have shown the existence of a universal zero conductivity condition in the complex frequency plane applicable to all materials. It sheds light on limits to absorption rate in plasmonic devices and can emerge as an important design criterion to optimize transduction rate in solar cells and single photon detectors.

\bibliography{references_UZCM.bib}


\appendix

\section{Proof of the zero conductivity condition}
We directly start from the Maxwell equations in frequency domain, 
\begin{align} \label{Maxwell}
\nabla \times E &=i\omega \mu_0 H, \nonumber  \\
\nabla \times H &=-i\omega \epsilon_0 \epsilon E. 
\end{align}
For a closed structure or a unit cell of a periodic structure\cite{Raman2013}, we have 
\begin{align}
0 &=\int (E \times H^*)\cdot dS \nonumber  \\
&=\int \nabla \cdot(E \times H^*) dV \nonumber \\
&=\int [H^* \cdot (\nabla \times E)-E \cdot (\nabla \times H^*)]dV,
\end{align}
where $S$ is a surface that encloses the whole structure of interest, $V$ is the corresponding volume. 
Substituting the Maxwell equation Eq.~\ref{Maxwell} in, one further get 
\begin{align}
0 &=\int [H^* \cdot (i\omega \mu_0 H)-E \cdot (-i\omega \epsilon_0 \epsilon E)^*]dV \nonumber \\
\Rightarrow & \int ( i\omega \mu_0 |H|^2-i\omega^* \epsilon_0 \epsilon^* |E|^2)dV=0  \nonumber \\
\Rightarrow & \int ( \omega \mu_0 |H|^2-\omega^* \epsilon_0 \epsilon^* |E|^2)dV=0  \nonumber \\
\Rightarrow & \int ( |\omega|^2 \mu_0 |H|^2-(\omega^*)^2 \epsilon_0 \epsilon^* |E|^2)dV=0  \nonumber \\
\Rightarrow & \int ( |\omega|^2 \mu_0 |H|^2-\omega^2 \epsilon_0 \epsilon |E|^2)dV=0  \nonumber \\
\end{align}
Now for every region $V_j$ inside the whole volume $V$ we define 
\begin{equation}
t_j=\frac{\int \epsilon_0 |E|^2 dV_j}{\int  |\omega|^2 \mu_0 |H|^2 dV}.
\end{equation}
Clearly $t_j$ are positive valued. We then find
\begin{equation}
\sum_j \omega^2 \epsilon_j t_j=1,
\end{equation}
or 
\begin{equation} \label{main}
\sum_j \epsilon_j t_j=\omega^{-2}.
\end{equation}
This simple equation is the starting point of the following argument. 

First note that $\omega$ should lie in the lower space of complex frequency (we use $e^{-i\omega t}$ time dependence), so $-\pi/2<arg(\omega)<0$. Therefore $0<arg(\omega ^{-2})<\pi$, which means $\omega^{-2}$ must have a positive imaginary part, that is, $\im(\omega ^{-2})>0$. This means the left hand side of Eq.~\ref{main} must also have a positive imaginary part, i.e., 
\begin{equation}
\sum_j \im(\epsilon_j) t_j>0. 
\end{equation}

As all $t_j$ are positive, the largest value of $\im(\epsilon_j)$ must be positive, or 
\begin{equation} \label{max}
\max_j(\im \epsilon_j(\omega))>0. 
\end{equation}
This equation suggests that for any eigenmode $\omega$, there must be at least one region/material whose permittivity has a positive imaginary part at this complex $\omega$.

For each region $j$ and given real part of frequency $\omega_R$, we can find $\omega_I^{j}$ such that
\begin{equation}
\im\epsilon_j(\omega_R-i\omega_I^{j})=0, 
\end{equation}
where $\omega_I^{j}$ should be positive valued. Eq.~\ref{max} then leads to a bound for the eigenmode $\omega=\omega_R-i\omega_I$,
\begin{equation}
\omega_I<\max_j \omega_I^{j}.
\end{equation}
Note that this equation exactly means that the material loss rate cannot go beyond the zero conductivity condition. 
Generally speaking, if we call a region effectively active/passive if its imaginary part of permittivity is negative/positive  at complex frequency, the eigen-frequency then cannot go beyond the point where all regions become effectively active. Or there must be at least one effectively passive region at the eigen-frequency. Furthermore, we expect this `principle' should apply to other generalized susceptibilities which relate to energy dissipation in various physical phenomena.

\section{Conductivity of Graphene}

The conductivity of graphene is given by \cite{wunsch2006dynamical,hwang2007dielectric} : 
\begin{align}
\sigma_{intra} &= \frac{2i}{\pi}\left(\frac{K_bT}{(\omega+i\tau^{-1})\hbar}\right)\ln\left( 2\cosh\left(\frac{E_F}{2T K_b}\right)\right)\\
\begin{split}
\sigma_{inter} &= \frac{1}{4}\left[\frac{1}{2} + \frac{1}{\pi}\tan^{-1}\left(\frac{\omega \hbar -2E_F}{2T K_b}\right) \right. \\ 
&-\left. \frac{i}{2\pi}\ln\left(\frac{\omega \hbar +2 E_F}{(\omega\hbar -2 E_F)^2 + (2 T K_B)^2}\right) \right]
\end{split}\\
\sigma&=\sigma_{intra}+\sigma_{inter}
\end{align}
Relaxation time $\tau=10^{-12}$, Temperature $T=300 K$. Where $\sigma_{intra}$ and $\sigma_{inter}$ corresponds to intera-band and inter-band conductivity, respectively; $K_B$ is the Boltzman constant, and $E_F$ is the Fermi Energy. 

\section{Definition of absorption, scattering and extinction cross-section}
\begin{eqnarray}
&C_{abs}&=\frac{P_{abs}}{\textrm{Incident~intensity}}\\
&C_{scattering}&=\frac{P_{scattered}}{\textrm{Incident~intensity}}\\
&C_{ext}&=C_{scattering}+C_{abs}
\end{eqnarray}
$C_{abs}$, $C_{scattering}$ and $C_{ext}$ corresponds to the absorption, scattering and extinction cross-sections, respectively. Cross-section is in units of area. For cylinder, it is computed per unit length. $P_{abs}$ and $P_{scattering}$ are the absorbed power and scattered power, respectively. 

If $A_m$ is the amplitude of the scattered field from the cylinder (normalized w.r.t. the incident field), then the scattering and extinction cross-section is given by \cite{bohren2008absorption},
\begin{equation}
C_{scattering}=\frac{4}{k_0}\left[\sum_m |A_m|^2\right]
\end{equation}
\begin{equation}
C_{extinction}=\frac{4}{k_0}Re\left[\sum_m A_m\right]
\end{equation}

\section{Eigenmodes of Nanowire}
A metallic nano-wire supports a plasomonic mode for transverse magnetic polarization. In the nano-wire, a mode loses its energy due to material loss in the metal as well as due the the radiation of the wave in the surrounding medium. The resonant frequencies of the eigen modes are the solution to the equation,
\begin{equation}
\sqrt{\epsilon_{0}}J_m^\prime\left(\sqrt{\epsilon_1} k a\right)H_m^{(1)\prime}\left(\epsilon_0 k a\right)-\sqrt{\epsilon_{1}}J_m\left(\sqrt{\epsilon_1} k a\right)H_m^{(1)\prime}\left(\epsilon_0 k a\right)=0
\label{eq_6}
\end{equation}
where $a$ is the radius of the nano-wire, $k=\omega/c$ is the phase constant in free space, $\epsilon_1$ is the complex relative dielectric constant of the nano-wire, $\epsilon_0$ is the relative dielectric constant of the surrounding medium, $J_m$ is the $m^{th}$ order Bessel function, and $H_m^{(1)}$ is the Hankel function of the $m^{th}$ order and first kind. Eq(\ref{eq_6}) has a solution only for complex values of $\omega$, where the imaginary component of the solution is a measure of the decay rate of the fields.%

\section{Optical absorption rate in Nanospheres}

\begin{figure}[t]
	\centering
	\subfigure[]{\includegraphics[width=0.45\linewidth]{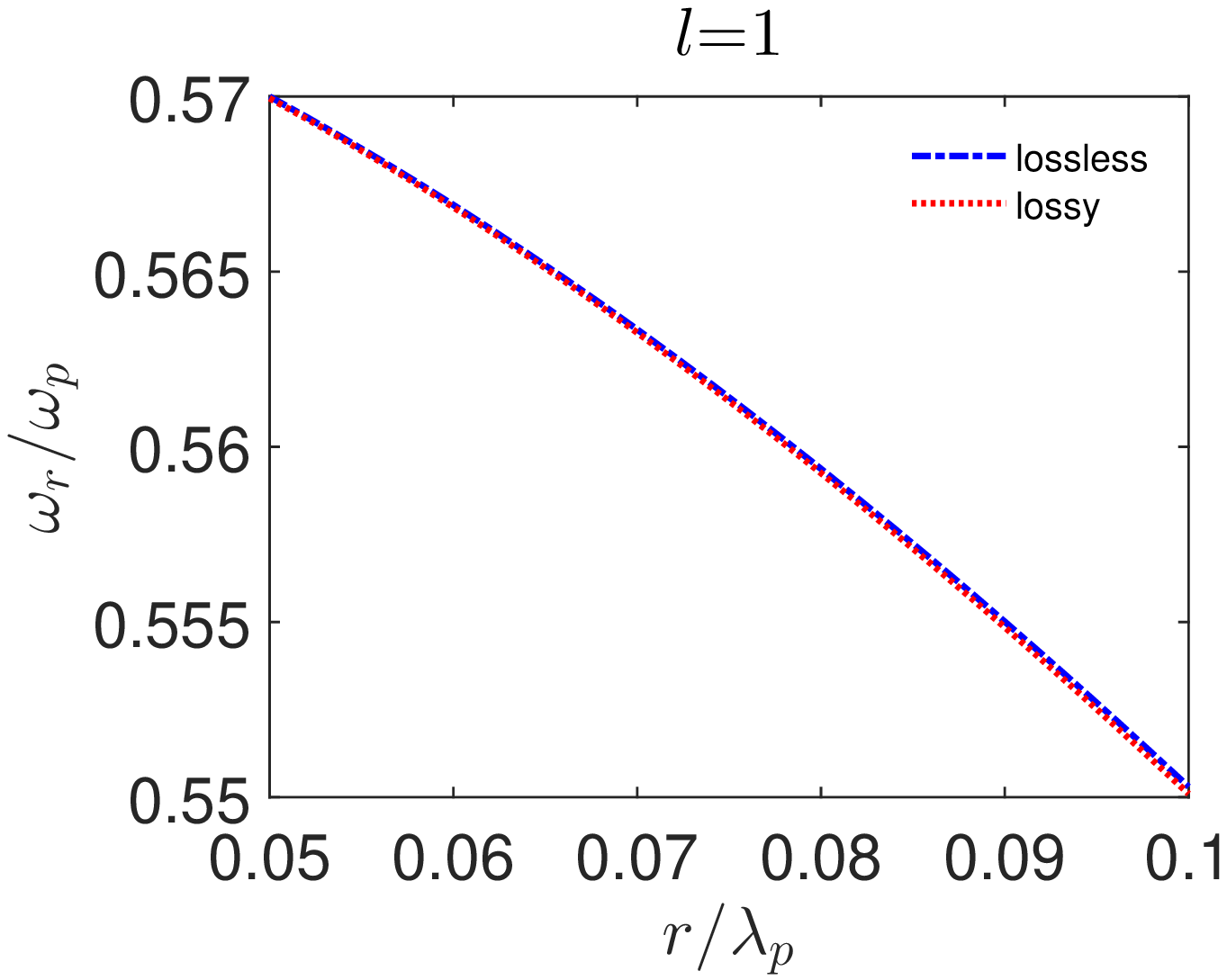}}
	\subfigure[]{\includegraphics[width=0.45\linewidth]{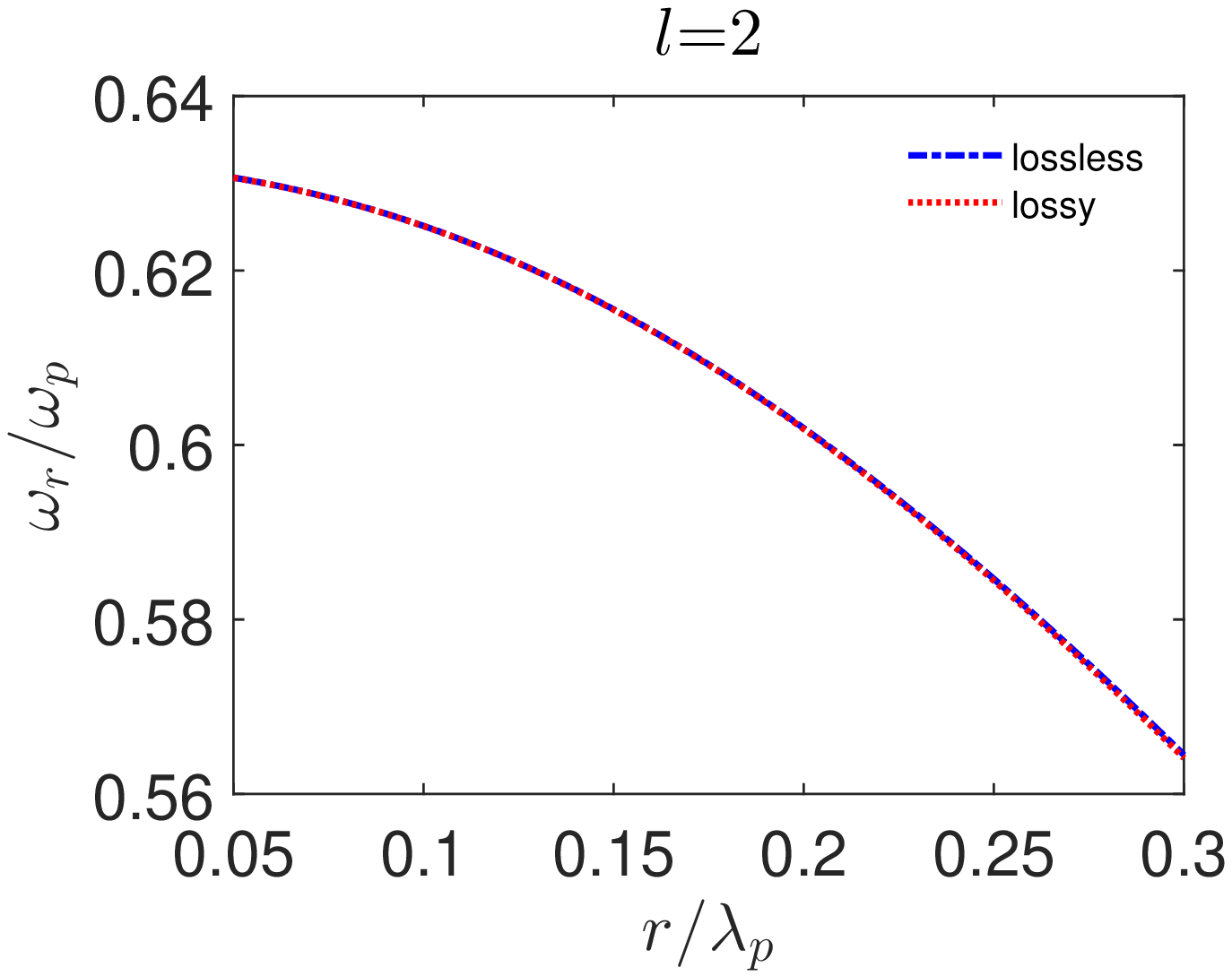}}\\
	\subfigure[]{\includegraphics[width=0.45\linewidth]{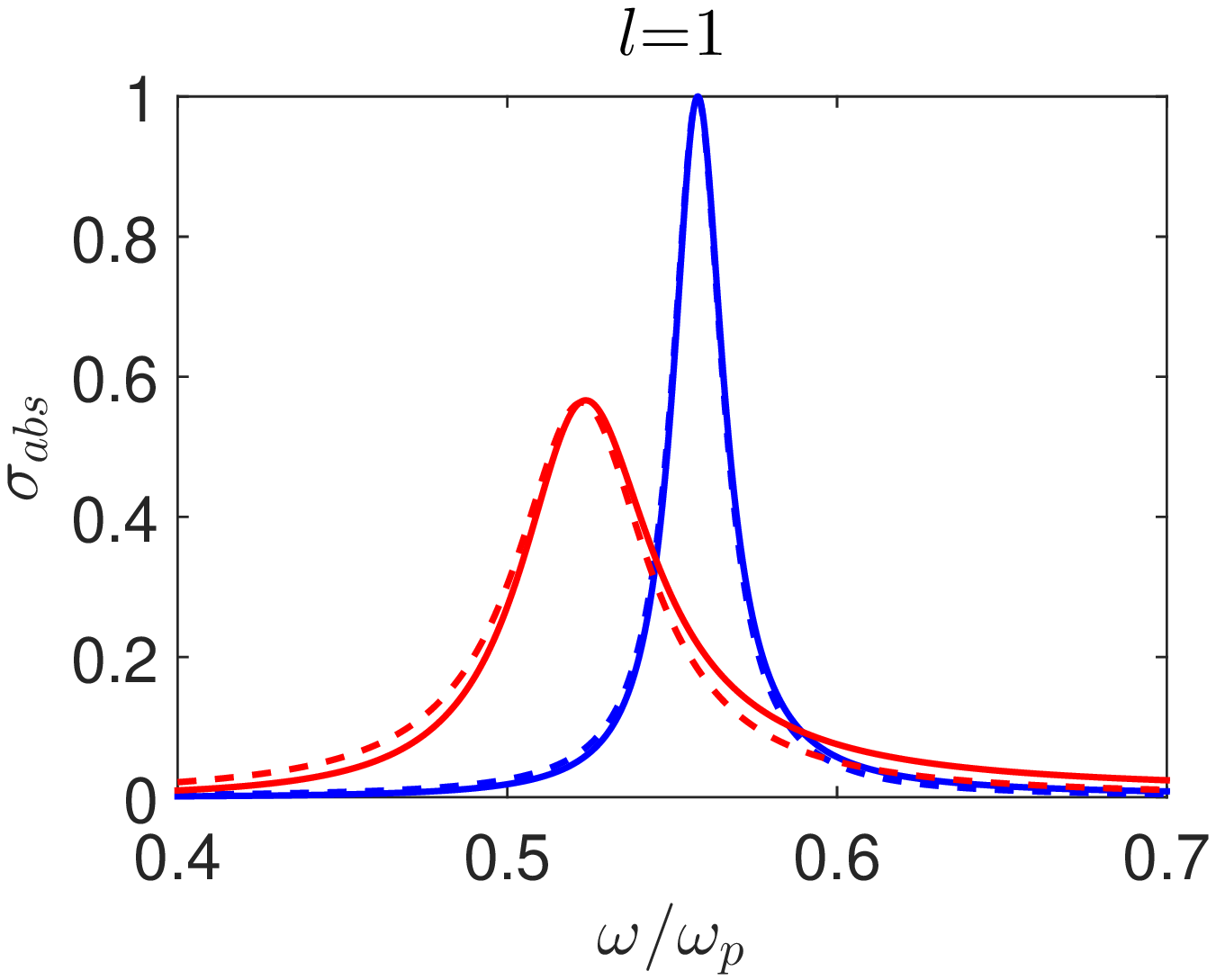}}
	\subfigure[]{\includegraphics[width=0.45\linewidth]{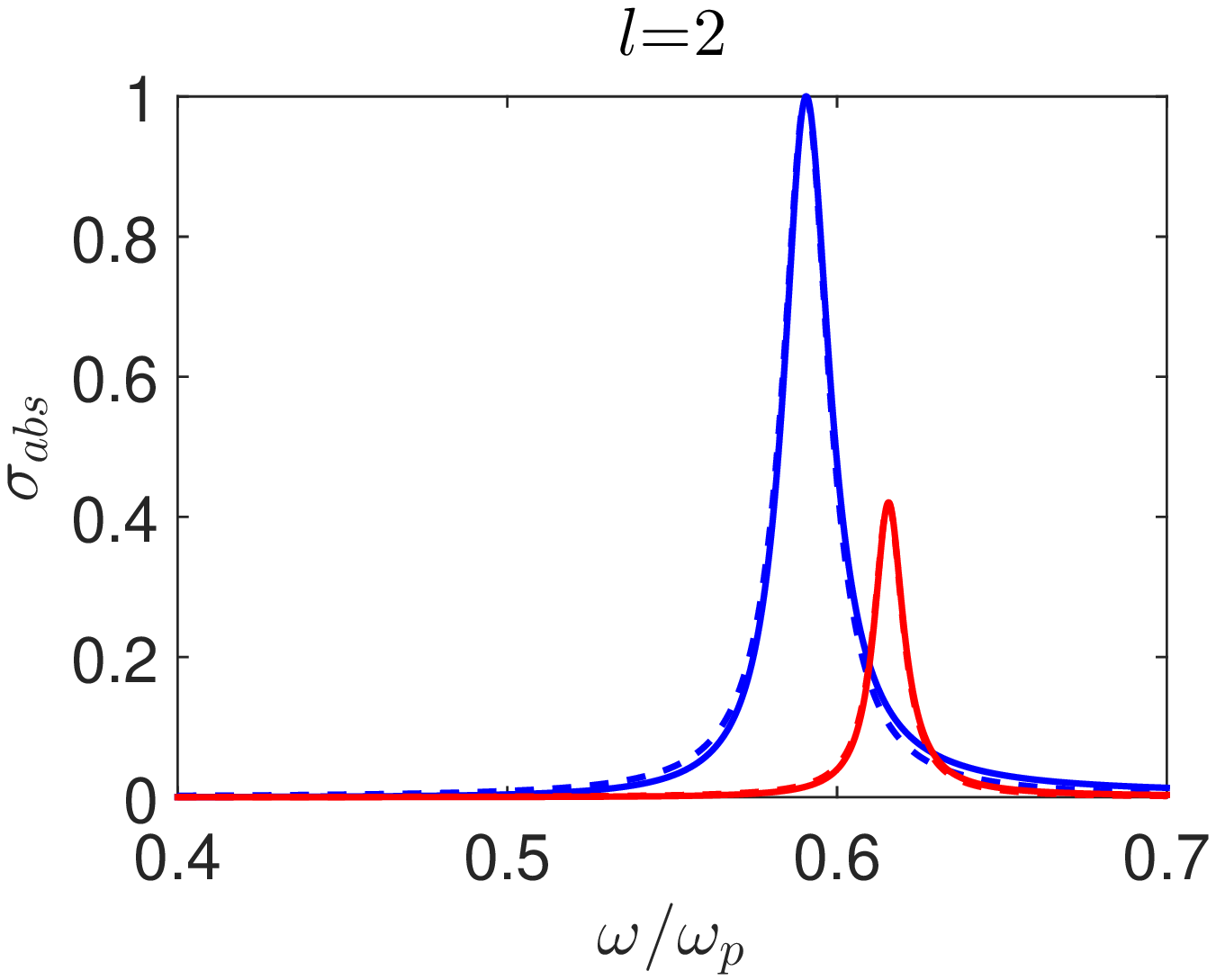}}
	\caption{\label{fig:sphere_wi_L1_supp} Comparison of the real component of eigen-frequency (resonant frequency) of surface plasmons in sphere for the lossless and lossy material, when (a) $l=1$, (b) $l=2$. It can be seen that the resonant frequency for lossy and lossless case are coinciding at all values of $r$. Therefore, material loss can be approximated by $\gamma=\gamma_t-\eta$. Lower panel shows normalized absorption cross section of a lossy sphere for (c) $l=1$ and (d) $l=2$, calculated by Mie theory(solid curve) and CMT(dash curve). (c)The blue curve represents a sphere radius of 0.084$\lambda_p$, where the maximum value of normalized absorption cross section is achieved; the red curve represents a sphere radius of 0.15$\lambda_p$ for comparison. (d) The blue curve represents a sphere radius of 0.234$\lambda_p$, where the maximum value of normalized absorption cross section is achieved. The red curve represents a sphere radius of 0.15$\lambda_p$ for comparison.}
\end{figure}

To compute the eigenfrequencies, we find the poles of the Mie scattering coefficients of TM modes, 
\begin{equation}
T_l = \frac{\epsilon j_l(k_mr)[k_0rj_l(k_0r)]'-j_l(k_0r)[k_mrj_l(k_mr)]'}{\epsilon j_l(k_mr)[k_0rh_l(k_0r)]'-h_l(k_0r)[k_mrj_l(k_mr)]'}
\end{equation}
here $k_0=\omega/c$ is the free space wavevector, $k_m=\sqrt{\epsilon} \omega/c$ is the wavevector in bulk medium. $j_l$ and $h_l$ are $l$-th order Bessel function of first and third kind, respectively. 

We compute the eigenfrequency for a lossless sphere to find the radiation loss and for a lossy sphere to find the total modal loss. 
Here $\eta$ is the modal loss when the material is lossless, that is, the radiation loss. $\gamma_t$ is the modal loss when the material is lossy, which includes the radiation loss and the material loss. We then approximate the material loss by $\gamma=\gamma_t-\eta$. This approximation is justified because the resonant frequency ($\omega_r$) for lossy and lossless nano-sphere are almost identical for all values of $r$, as shown in  Fig.~\ref{fig:sphere_wi_L1_supp}(a)-(b).

The absorption cross section due to this $l$-th order TM mode is given by
\begin{equation}
\sigma_{Mie} = \frac{\lambda^2}{8\pi}(2l+1)(4Re(T_l)-4|T_l|^2)
\end{equation}

In the coupled mode theory(CMT), the absorption cross section for a lossy sphere which support a single resonance is \cite{ruan2011design}
\begin{equation}
\sigma_{CMT} = \frac{\lambda^2}{8\pi}(2l+1)\frac{4\eta \gamma}{(\omega-\omega_0)^2+(\eta+\gamma)^2}
\end{equation}
$\eta$ is the radiation loss, $\gamma$ is the material loss. The absorption cross section is normalized by the factor $\frac{\lambda^2}{8\pi}(2l+1)$. From CMT, it is clear that the maximum normalized absorption cross section is one, which is achieved at the critical coupling condition when the structure is excited at the resonant frequency $\omega_0$, and the radiation loss late equals the material loss rate. This condition is equivalent to the design of coherent perfect absorbers which controls the zeros of the scattering matrix. Note, the CPA does not provide insight into the optimal absorption rate which is provided by our theory of the zero conductivity condition. The absorption cross-section as function of frequency at radius at which absorption cross-section is maximized in shown in Fig.~\ref{fig:sphere_wi_L1_supp}(c)-(d).


\end{document}